\documentclass[twocolumn]{aastex61}
\usepackage{natbib}
\usepackage{color}
\usepackage{amsmath}
\usepackage{soul}

\def\kms{{\rm km s}^{-1}}

\def\Myr{{\rm\thinspace Myr}}
\newcommand{\lamR}{\mbox{$\lambda_{\rm R}$}}

\received{July 1, 2016}
\revised{September 27, 2016}
\accepted{\today}
\submitjournal{ApJ}
\shorttitle{early-type galaxy spin evolution}
\shortauthors{Choi et al.}

\begin{document}

\title{Early-type galaxy spin evolution in the Horizon-AGN simulation}

\correspondingauthor{Hoseung Choi}
\email{choi.h@yonsei.ac.kr, yi@yonsei.ac.kr}

\author[0000-0001-7229-0033]{Hoseung Choi}
\affiliation{Department of Astronomy and Yonsei University Observatory, \\
Yonsei University, Seoul 03722, Republic of Korea}

\author[0000-0002-4556-2619]{Sukyoung K. Yi}
\affiliation{Department of Astronomy and Yonsei University Observatory, \\
Yonsei University, Seoul 03722, Republic of Korea}

\author{Yohan Dubois}
\affiliation{Institut d'Astrophysique de Paris, Sorbonne Universit\'{e}s, UMPC Univ Paris 06 et CNRS, UMP 7095, 98 bis bd Arago, 75014 Paris, France}

\author[0000-0002-3950-3997]{Taysun Kimm}
\affiliation{Department of Astronomy and Yonsei University Observatory, \\
Yonsei University, Seoul 03722, Republic of Korea}

\author{Julien. E. G. Devriendt}
\affiliation{Dept of Physics, University of Oxford, Keble Road, Oxford OX1 3RH UK}
\affiliation{Université de Lyon, Université Lyon 1, ENS de Lyon, CNRS, Centre de Recherche Astrophysique de Lyon UMR5574, F-69230 Saint-Genis-Laval, France}

\author{Christophe Pichon}
\affiliation{Institut d'Astrophysique de Paris, Sorbonne Universit\'{e}s, UMPC Univ Paris 06 et CNRS, UMP 7095, 98 bis bd Arago, 75014 Paris, France}
\affiliation{Korea Institute of Advanced Studies (KIAS) 85 Hoegiro, Dongdaemun-gu, Seoul, 02455, Republic of Korea}

\begin{abstract}
Using the Horizon-AGN simulation data, we study the relative role of mergers and environmental effects in shaping the spin of early-type galaxies (ETGs) after $z \simeq 1$. 
We follow the spin evolution of 10,037 color-selected ETGs more massive than 10$^{10} \rm \, M_{\odot}$ that are divided into four groups: cluster centrals (3\%), cluster satellites (33\%), group centrals (5\%), and field ETGs (59\%). 
We find a strong mass dependence of the slow rotator fraction, $f_{\rm SR}$, and the mean spin of massive ETGs.
Although we do not find a clear environmental dependence of $f_{\rm SR}$, a weak trend is seen in the mean value of spin parameter driven by the satellite ETGs as they gradually lose their spin as their environment becomes denser.
Galaxy mergers appear to be the main cause of total spin changes in 94\% of central ETGs of halos with $M_{vir} > 10^{12.5}\rm M_{\odot}$, but only 22\% of satellite and field ETGs.
We find that non-merger induced tidal perturbations better correlate with the galaxy spin-down in satellite ETGs than mergers.
Given that the majority of ETGs are not central in dense environments, 
we conclude that non-merger tidal perturbation effects played a key role for the spin evolution of ETGs observed in the local ($z < 1$) universe.
\end{abstract}


\section{Introduction} \label{sec:intro}
Massive early-type galaxies (ETGs) are believed to form through numerous mergers, and are preferentially found in high-density environments \citep[the so-called morphology-density relation,][]{Dressler1980}.
The importance of mergers in shaping the morphology of galaxies has been well recognized for decades \citep[e.g.,][]{Toomre1972,Barnes1988,DiMatteo2007,Dekel2009,Naab2014}.
It was the introduction of integral field unit spectroscopic (IFU) surveys that boosted the investigations on the link between mergers and the properties of their remnants.
Notably, the differential rotation of early-type galaxies observed with IFU surveys has drawn much attention \citep{Emsellem2007,Cappellari2011}.
Many theoretical studies immediately followed with scenarios addressing the wide range of observed ETG spin with a focus on the role of mergers. 

\citet{Bois2011} demonstrated that binary merger remnants produced under various merger conditions can reproduce the observed ETG spin-ellipticity distribution \citep{Emsellem2007}.
By conducting a set of 44 zoom-in simulations of central galaxies, \citet{Naab2014} claimed that the main driver for the different types of galaxy spins is the difference in their merger histories.
\citet{Khochfar2011} took a semi-analytic approach and concluded that massive ETGs that grow predominantly through mergers tend to evolve into slow rotators (SRs).

Studies based on cosmological simulations have shown correlations between galactic spin, ellipticity, and stellar mass, which compare reasonably well to various observations.
\citet{Lagos2017a, Lagos2017b} analyzed the EAGLE simulation and found that wet mergers can spin-up galaxies, whereas dry mergers tend do the opposite. 
Other merger properties such as the merger orbit or the galaxy spin alignment have been found to be only marginally relevant.
Based on an analysis of the Illustris simulation data, on the other hand, \citet{Penoyre2017} concluded that it is the re-accretion of cold gas that promotes the galactic spin-up, regardless of whether a galaxy undergoes a wet or dry merger.
Despite some disparity in the details of such scenarios, it is clear that frequent mergers result in SRs, particularly at low redshifts ($z < 2$), where ETGs have little chance of accreting cold gas.

Because both the stellar mass growth and spin-down of ETGs are driven by mergers, it is reasonable to expect a correlation between stellar mass and kinematic properties.
In accordance with the scenarios mentioned above, multiple studies have reported a clear dependence of galactic spin with the stellar mass.
In the stellar mass range above $\sim 5\times 10^{10}M_{\odot}$, the fraction of slow rotators ($f_{\rm SR}$) increases with mass from $10\% - 20\%$ all the way up to over 80\% at $\sim10^{11.7} \thinspace \rm M_{\odot}$ \citep{VandeSande2017b, Veale2017a, Brough2017a, Greene2017a}.

Mergers being one of many environmental effects, it is important to understand the relative significance of galaxy mass and local environment.
Motivated by the classic morphology-density relation,
\citet{Cappellari2011, Cappellari2016} suggested a kinematic morphology-density relation which quantify how ETGs in high-density environments have a  lower spin than ETGs in low-density environments.
The kinematic morphology-density relation has been supported by the observations of a handful of clusters \citep{DEugenio2013, Houghton2012, Jimmy2013, Scott2014, Fogarty2014}.
The dependence of galaxy kinematics on the environment, however, is being challenged.
Compiling observations of 22 central galaxies from the literature,
\citet{Oliva-altamirano2017} showed that the dependency of central galaxy spin on the host halo mass is very weak at 1.8 $\sigma$.
Moreover, recent studies based on more extensive samples of IFU surveys have also claimed that there is only a negligible environmental dependency at a fixed stellar mass, suggesting that the apparent kinematic morphology-density relation is simply driven by the mass dependence of ETGs on the environment \citep{Brough2017a, Veale2017a, Greene2017a, Greene2017b}.

Nevertheless, the role of environment may still be important to understand the evolution of ETGs in a wider scope.
Unlike massive ETGs, a clear environmental dependence was reported for dwarf ETGs with $\rm M_{\ast} < 5 \times 10^{9}\rm \thinspace M_{\odot}$ in the Virgo cluster \citep{Guerou2015, Toloba2015}.
It is not surprising that the kinematic transformation of dwarf ETGs is driven through the environmental effects of the large host potential because smaller galaxies are more easily affected by their environment \citep[][c.f. \citealt{Janz2017}]{Vollmer2009, Boselli2009, Boselli2014}.
The relative importance of mergers and environmental effects in the mass range between massive ETGs and dwarf ETGs has been discussed in a numerical study by \citet{Choi2017}.
Based on 16 zoom-in hydrodynamic simulations of galaxy clusters, they found that 
galaxy mergers are the main driver of spin evolution for the most massive ETGs ($M_{\ast} > 5 \times 10^{10} \rm \thinspace M_{\odot}$), while they cannot fully account for the spin change of intermediate mass systems ($5 \times 10^{9} \rm \thinspace M_{\odot} < \rm M_\ast < 5 \times 10^{10} \rm \thinspace M_{\odot}$).
While the spin of most massive galaxies is thought to be driven by numerous mergers,
more investigations are called for on the spin evolution of dwarf and intermediate-mass ETGs and their connections to mergers and environments.

In the present study, we investigated the spin properties of a large number of intermediate to high mass ETGs based on the cosmological simulation, Horizon-AGN.
We show that
1) mergers are important in shaping ETG spin, but only for massive central galaxies.
2) satellite galaxies do have lower spins compared to field counterparts,
3) and it is due to tidal perturbation in high-density environments, not mergers.
In Section \ref{sec:methods}, we describe the simulation data, post-processing, and derivation of the galaxy properties.
The relation between spin and other galaxy properties in various subsamples is given in Section \ref{sec:Results1}.
Finally, we summarize and discuss the results in Section \ref{sec:summary}.

\section{Methods} \label{sec:methods}
\subsection{Horizon-AGN Simulation}
We used the Horizon-AGN simulation \citep{Dubois2014}, which was performed using the adaptive mesh refinement code, {\sc ramses} \citep{Teyssier2002}.
The assumed cosmology in the simulation is a flat $\Lambda$CDM universe with a Hubble constant of $H_{0} = 70.4 \, \kms$ Mpc$^{-1}$, a baryon density of $\Omega_{b} = 0.0456$,
a total matter density of $\Omega_{m} = 0.272$, a dark energy density of $\Omega_{\Lambda} = 0.728$, 
an rms fluctuation amplitude at $8\,h^{-1}\,{\rm Mpc}$ of $\sigma_{8} = 0.809$, and a spectral index $n = 0.963$, which is
consistent with the seven-year Wilkinson Microwave Anisotropy Probe (WMAP) analysis \citep{Komatsu2011}.
The simulation volume is $(100\, h^{-1} {\rm Mpc})^3$,
and contains 14 clusters of mass $M_{\rm vir} > 10^{14} \, \rm M_{\odot}$,
367 groups of mass $ 10^{13}\, M_{\odot}  < M_{\rm vir} < 10^{14}\, M_{\odot} $, and field ETGs.

Radiative gas cooling is modeled following \citet{Sutherland1993}. 
A uniform UV background heating is activated after $z_{\rm reion} = 10$ based on \citet{Haardt1996}. 
Cells with hydrogen number density above $n_{0} = 0.1 \,{\rm H \, cm}^{-3}$ are allowed to form stellar particles through a Poisson random process \citep{Rasera2006, Dubois2008}. 
A 2\% star formation efficiency per free-fall time is assumed in the Schmidt law $\dot{\rho} = \varepsilon_{\ast}\rho_{g}/t_{\rm ff}$, where $\rho_{\rm g}$ is the gas density, and $t_{\rm ff}=\sqrt{3\pi/(32 G \rho_{\rm g})}$ is the local free-fall time \citep{Kennicutt1998}.
Stellar feedback from supernova Type Ia, II, and stellar winds is implemented assuming a Salpeter \citep{Salpeter1955} initial mass function \citep{Kaviraj2017}.
The black hole growth is modeled by the Bondi-Hoyle-Lyttleton accretion \citep{Hoyle1939, Bondi1944}.
When the gas accretion rate is low, the black hole launches bi-polar jets (radio mode),
whereas active galactic nuclei deposit thermal energy isotropically when the accretion rate is high (quasar mode) \citep{Dubois2012}.

\subsection{Galaxy identification and merger tree} \label{sec:galaxydetection}
The galaxies in the simulation were identified using HaloMaker through the AdaptaHOP method \citep{Aubert2004},
with the most massive sub-node mode \citep{Tweed2009} applied for stellar particles.
A minimum of 64 stellar particles, or $M_{*} = 5 \times 10^{7} \, \rm M_{\odot}$, were used to define a galaxy,
but we limit our spin analysis to galaxies more massive than $M_{*} = 10^{10} \thinspace \rm M_{\odot}$ to ensure robust measurements.
The smaller galaxies identified are only considered as merging satellites or sources of tidal perturbation in the later part of the analysis.

To determine the size of galaxies, we first measured the stellar mass above 
the surface density cut ($\Sigma_{\rm M} > 10^{6} \, \rm M_{\odot}\, kpc^{-2}$), which is used to compute
a tentative half-mass radius ($R_{\rm eff,ten}$). The threshold is chosen  
so that it reasonably covers the entire regions of isolated galaxies.
We then re-calculated the half-mass radius ($R_{\rm eff}$) using the star particles within $R\le 4\, R_{\rm eff,ten}$, 
and determined the total stellar mass by integrating the mass of star particles within $R\le 4\, R_{\rm eff}$.

To follow the evolution of galaxy spin, we constructed galaxy merger trees using 787 snapshots of the Horizon-AGN simulation. 
The time interval between snapshots  corresponds to $\Delta a_{exp} \approx 0.0001$, or $17\Myr$. 
Note that the number of snapshots is larger than usual for cosmological simulation datasets, allowing us to precisely determine the progenitor-descendant relation, and monitor the beginning and end of the mergers. 

We defined the beginning of a merger as when a satellite galaxy crosses three times the sum of the radii of the host and satellite galaxy.
We also required the orbital angular momentum of the satellite galaxy to stay below the initial value at three times the sum of the radii.
If the satellite re-acquires the orbital angular momentum or moves outside the radial distance limit, the merger is considered to exert no effect on the host galaxy in the meantime.
The end of a merger is defined as the time at which the satellite is no longer detected by HaloMaker.
By filtering out fly-by-like interactions in the very early stages of minor mergers, we determined the merger stages where the stellar component of the host galaxy actually feels disturbances.
All mergers above a merger mass ratio of 1:50 are considered, and mergers with a mass ratio larger than 1:4 are considered as major mergers.

\begin{figure}[t]
\figurenum{1}
\includegraphics[width=0.45\textwidth]{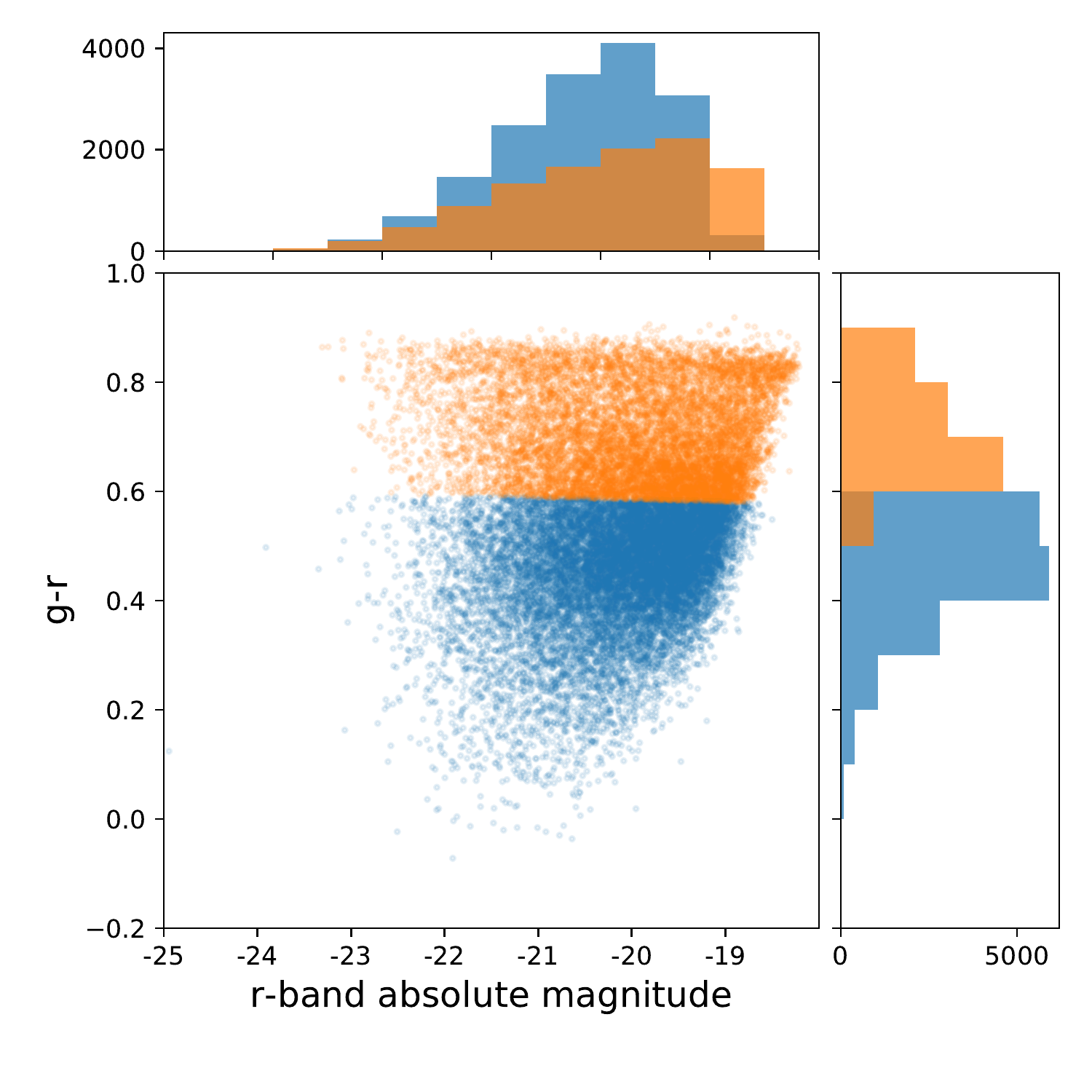}
\caption{Classification of early-type galaxies based on the color-magnitude diagram \citep{Gabor2010}. 
Orange dots indicate early-type galaxies, while late types are shown as blue.
Note that dust attenuation is not taken into consideration during the sample selection procedure to avoid contamination 
by dust-reddened late-type galaxies.
\label{fig:sampling}}
\end{figure}

\subsection{Spin parameter} \label{sec:lambda}
To compute the galactic spin parameter \citep[$\lambda_{\rm R}$, ][]{Emsellem2007} of simulated galaxies, 
we first generated two-dimensional projection maps of stellar luminosity, the luminosity-weighted line-of-sight velocity, and luminosity-weighted velocity dispersion along the $z$-direction of the simulation.
To ensure statistical reliability, we split each stellar particle into 60 pseudo particles, which were distributed following a Gaussian kernel with a standard deviation of 0.3 kpc.
Pseudo particles were grouped using the Voronoi binning technique \citep{Cappellari2003} to achieve a uniform statistical significance over the measurement points.
We used the publicly available MGE package \citep{Cappellari2002, Emsellem1994} to determine 
the optical center of a galaxy and measure the ellipticity $\varepsilon\, (\equiv 1-b/a)$, 
where $a$ and $b$ are the semi-major and semi-minor axis of the fitted ellipse, respectively.
We adopted the ellipticity $\varepsilon$ at $\sqrt{ab} \sim R_{\rm eff}$ as the representative $\varepsilon$ of a galaxy throughout this paper.
The center of velocity is computed by the mean velocity of the 10\% closest points to the optical center.

\begin{figure*}
\figurenum{2}
\includegraphics[width=0.95\textwidth]{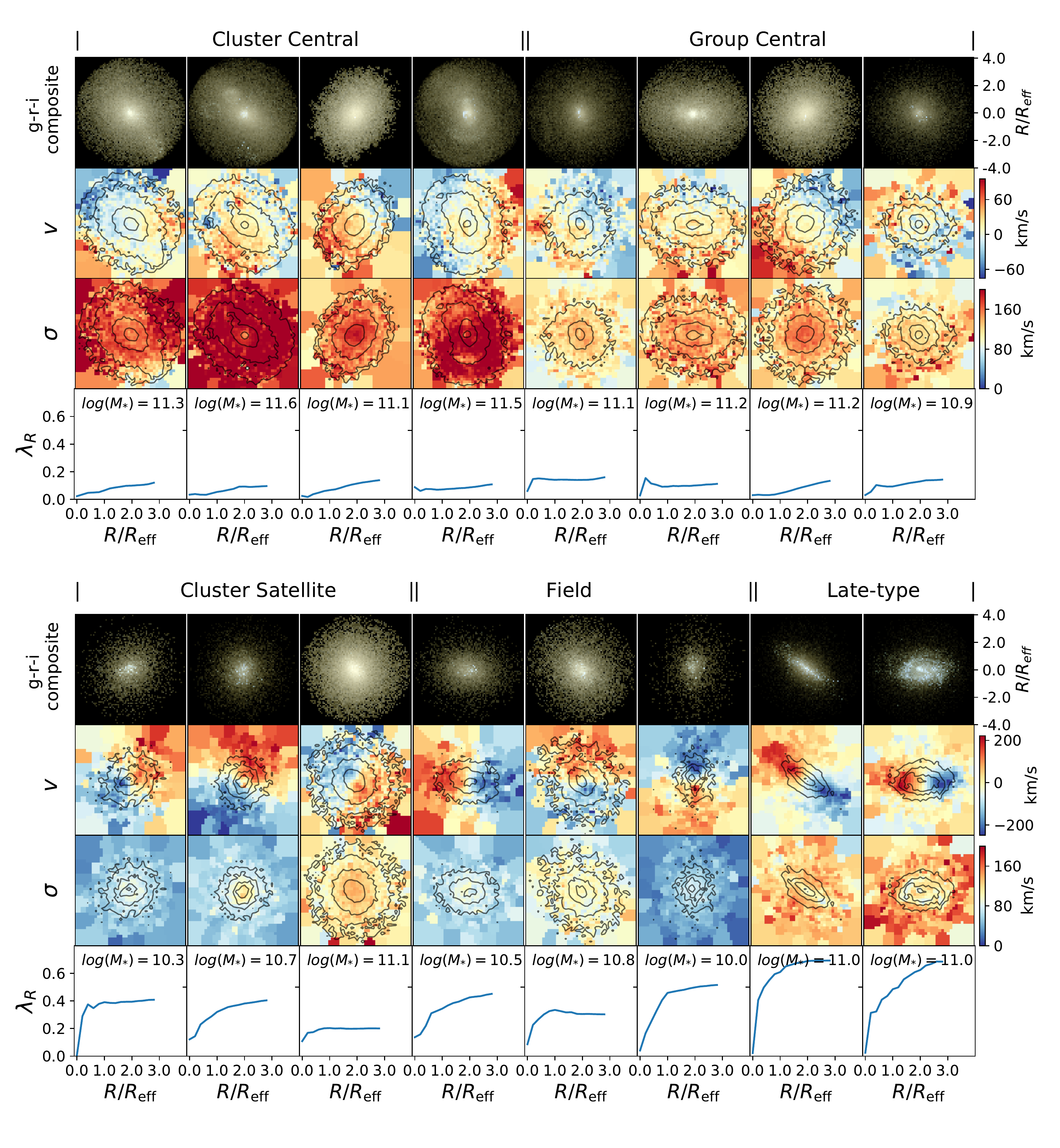}
\caption{Examples of 14 ETGs and two late-type galaxies from Horizon-AGN. 
From the top to bottom, each row shows the composite image of the SDSS $g$, $r$, and $i$ bands, 
line-of-sight velocity, line-of-sight velocity dispersion, and radial profile of $\lambda_{\rm R}$ of each galaxy. 
The last two late-type galaxies are not included in our ETG sample, but presented for comparison purposes. 
\label{fig:postage}}
\end{figure*}

The spin parameter of a galaxy was measured following the definition of \citet{Emsellem2007} as
\begin{equation}
\lambda_R=\frac{\Sigma_{i} F_i R_i
  |V_i|}{\displaystyle\Sigma_{i}F_i R_i \sqrt{V_i^2 +\sigma_i^2}}, 
\end{equation}
where $R_i$ is the radius of the concentric ellipse, $F_i$ is the dust-attenuated flux,
$V_i$ is the luminosity-weighted mean line-of-sight velocity of stellar particles,
and $\sigma_i$ is the line-of-sight velocity dispersion of the $i$-th spatial bin.
Note that we adopted the spin parameter measured within the fitted ellipse of $x^2/a^2 + y^2/b^2 = 1$ 
for a fair comparison with observations.
Figure \ref{fig:postage} shows the examples of the spin measurement for randomly selected galaxies of different 
morphology and environments (see Section~\ref{sec:subgroup}).

\subsection{Early-type galaxy sample}\label{sec:sampling}

We defined ETGs to be red galaxies\footnote{Another possible way to classify the morphology of 
simulated galaxies is based on the orbital eccentricity of the stellar particles and their direction with 
respect to the galaxy spin \citep{Abadi2003, Scannapieco2009, Sales2012}. However, relating 
kinematic classification to observational classification has ambiguities of its own \citep{Bottrell2017}.
Thus, we took a simpler approach based on the color-magnitude diagram.} following \citet{Gabor2010}, as
\begin{equation}
g-r > -0.0042 \, M_{\rm r} + 0.5,
\end{equation}
where $M_{\rm r}$ is the SDSS $r$-band absolute magnitude.
We calculated the galaxy's $g-r$ color by generating mock images using the stellar population model 
of \citet{Bruzual2003} assuming a Salpeter initial mass function. 
Note that dust extinction is not taken into account during the sample selection procedure to avoid possible contamination 
by dust-reddened late-type galaxies. 
This results in 10,037 ETGs out of the total 21,486 galaxies above $M_{\ast} = 10^{10} \, M_{\odot}$ ($\sim 45\%$) (Figure \ref{fig:sampling}). 
The fraction of ETGs is comparable to observations \citep[for example]{Skibba2009, Khim2015}.
As a sanity check, we also performed visual inspection of randomly selected subsample of 10\% of the total Horizon-AGN galaxies, 
and found that late-type galaxies are very rarely included in our ETG sample.
We also checked that using a different selection criterion \citep{Schawinski2014} does not change our main results.

Figure \ref{fig:postage} shows the examples of the spin measurement for randomly selected galaxies of different 
morphology and environments (see Section~\ref{sec:subgroup}). 
Also included is the spin parameter as a function of radius. 

\begin{figure*}
\figurenum{3}
\includegraphics[width=0.95\textwidth]{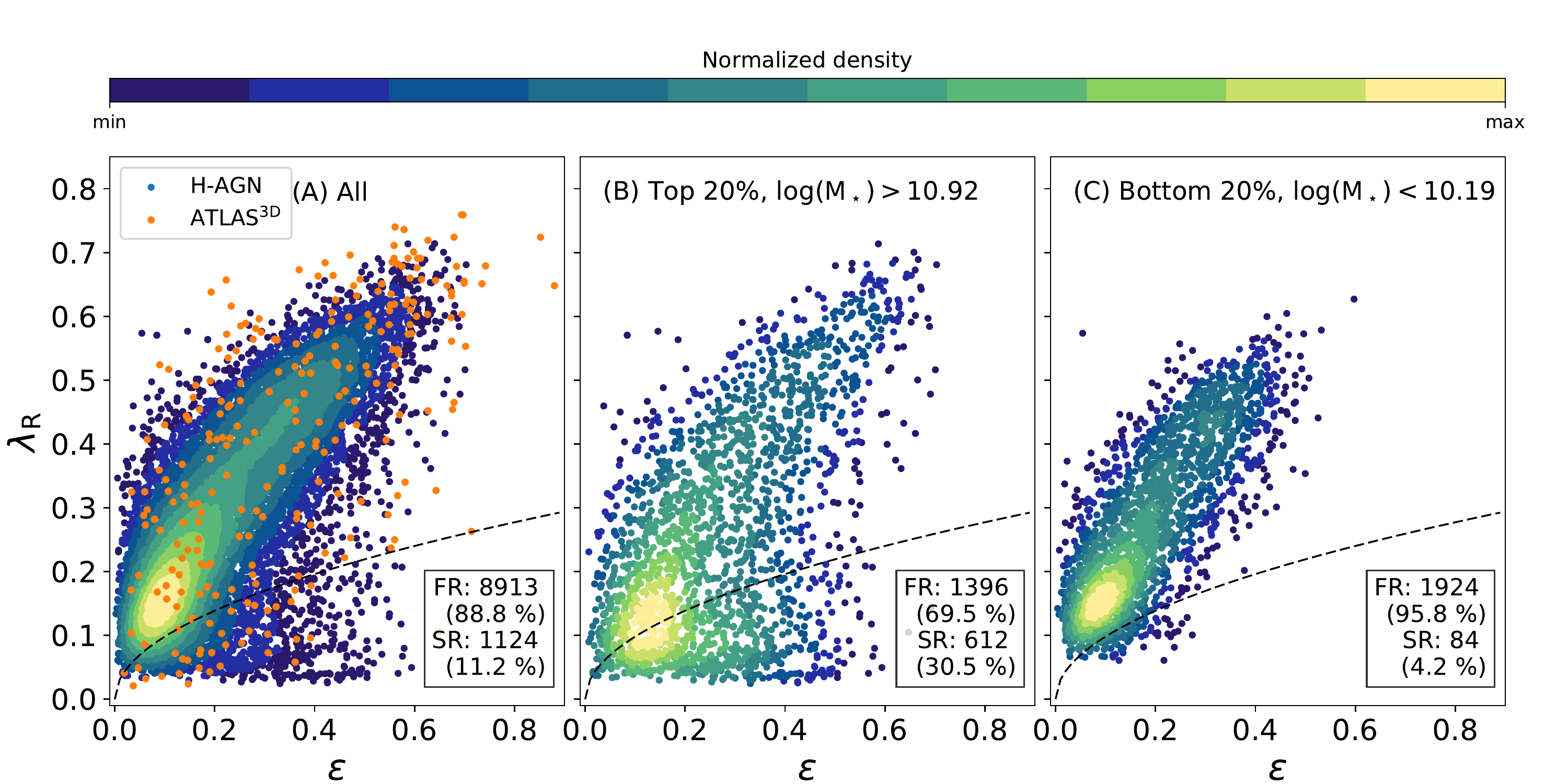}
\caption{Spin parameter \lamR\ versus ellipticity $\varepsilon$ for all sample ETGs (panel A, left). 
Orange points indicates 260 ETGs from ATLAS$^{\rm 3D}$.
The 10-step shades represent a kernel density estimation of the galaxy distribution normalized in each panel.
Slow rotators are defined by the demarcation line at $R_{e}$ from \citet{Emsellem2011}.
The middle panel shows the distribution for the top 20\% in mass ($M_{*} > 8.1 \times 10^{10} \thinspace \rm M_{\odot}$) (panel B), while the bottom 20\% in mass ($M_{*} < 1.5 \times 10^{10} \thinspace \rm M_{\odot}$) is shown in the right panel (panel C). 
\label{fig:l_e_mass}}
\end{figure*}

\section{Results}\label{sec:Results1}

\subsection{Mass and environmental dependence of the whole sample}\label{sec:mass_env_whole}

\begin{figure*}
\figurenum{4}
\includegraphics[width=0.9\textwidth]{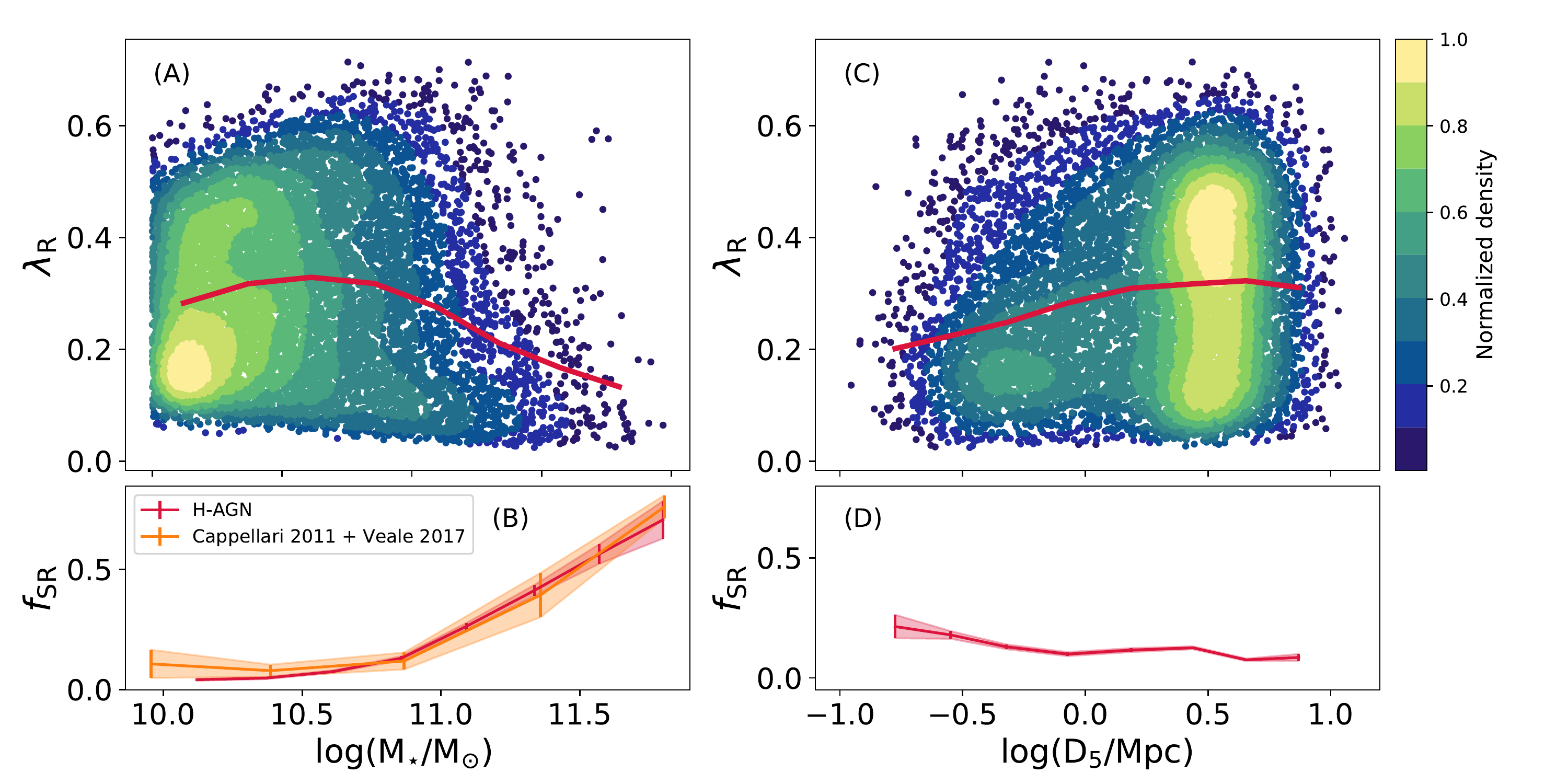}
\caption{Comparison of \lamR\ dependence on mass (left) and environment (right).
The upper panels show the distribution of \lamR\  with galaxy stellar mass. 
The 10-step shades represent a kernel density estimation of the galaxy distribution.
The red solid line denote the mean spin parameter measured.
In the lower left panel, we present the dependence of the fraction of slow rotators on stellar mass, which is largely consistent with observations \citep{Oliva-altamirano2017, Veale2017a, VandeSande2017b, Greene2017a}. 
It can be seen that most of the massive galaxies are concentrated in a low \lamR\ region, 
whereas the scatter increases significantly for less massive ETGs.
Compared to the mass dependency of $f_{\rm SR}$, the environmental dependency of $f_{\rm SR}$ is only marginal.
Nonetheless, the mean \lamR\ of the galaxies in very dense regions (D$_{5} < 1$Mpc) shows a weakly positive correlation with D$_{5}$.
\label{fig:l_e_mass_env}}
\end{figure*}

The distribution of spin parameter $\lambda_{\rm R}$ of the 10,037 ETGs is compared to the ATLAS$^{\rm 3D}$ 
observations \citep{Emsellem2011} in  Figure \ref{fig:l_e_mass}A.
The overall shapes of the two distributions are qualitatively similar.
With a sufficient number of galaxies, the simulated galaxy distribution appears smooth and displays a single peak at \lamR\ $\approx$ 0.15 and $\varepsilon$  $\approx$ 0.1. 
If the demarcation $\lambda_{R} = 0.31 \sqrt{\epsilon}$ of \citet[][dashed black line]{Emsellem2011} is used, the slow rotator fraction $f_{\rm SR}$ is 11.2\% (1124/10,037);
a value that is comparable to $14\% \pm{2}$ according to ATLAS$^{\rm 3D}$ \citep{Emsellem2011}.

The top and bottom 20\% of the sample galaxy based on their stellar mass are presented in Figure \ref{fig:l_e_mass}B and C , respectively.
A substantial fraction (30.5\%) of massive ETGs were found to be SRs, whereas only 4.2\% of low-mass ETGs were found to be so.
The range of $\varepsilon$ and \lamR\ values represented in Figure \ref{fig:l_e_mass}B and C is different.
Massive ETGs form a broad peak at $\varepsilon \approx 0.1, \lambda_{\rm R} \approx 0.1$ with notable scatters in both $\varepsilon$ and \lamR\ axes.
By contrast, low-mass ETGs are clustered tightly to form an elongated peak from $\varepsilon \approx 0.1, \lambda_{\rm R} \approx 0.2$ to $\varepsilon \approx 0.2, \lambda_{\rm R} \approx 0.45$.
Some slow rotators were found to have very high ellipticity with complex kinematic structures inside $1\,R_{\rm eff}$ owing to satellite remnants (fourth galaxy in Figure \ref{fig:postage}, for example); a result corroborated by \citet{Naab2014, Cappellari2016, VandeSande2017a}.

The correlation between stellar mass and $f_{\rm SR}$ is clearly visible, i.e., massive galaxies are preferentially slow rotators (Figure \ref{fig:l_e_mass_env}A).
However, the trend in mean \lamR\ against stellar mass is weaker than $f_{\rm SR}$ owing to the large scatter in \lamR\ (Figure \ref{fig:l_e_mass_env}B).
The scatters are markedly larger in low stellar mass bins compared to high stellar mass bins, suggesting that the same process responsible for a galaxy spin-down may also be responsible for a mass growth of massive galaxies.
Alternatively, the significant scatters in \lamR\ of low-mass bins imply that processes unrelated to mass growth are operating to shape galactic spin.

Next, we examined the relationship between \lamR\ and the galaxy's local environmental density D$_{5}$, which is defined as the distance to the fifth closest galaxy more massive than $10^{10}\, \rm M_{\odot}$.
The trends in both the \lamR\ distribution (Figure \ref{fig:l_e_mass_env}C) and $f_{\rm SR}$ (Figure \ref{fig:l_e_mass_env}D) are not as pronounced compared to the mass dependency.
Only a weak dependence of mean \lamR\ on the local density below D$_{5} \lesssim 3$Mpc was found, and no trend is evident in $f_{\rm SR}$.

\begin{figure}[h!]
\figurenum{5}
\includegraphics[width=0.5\textwidth]{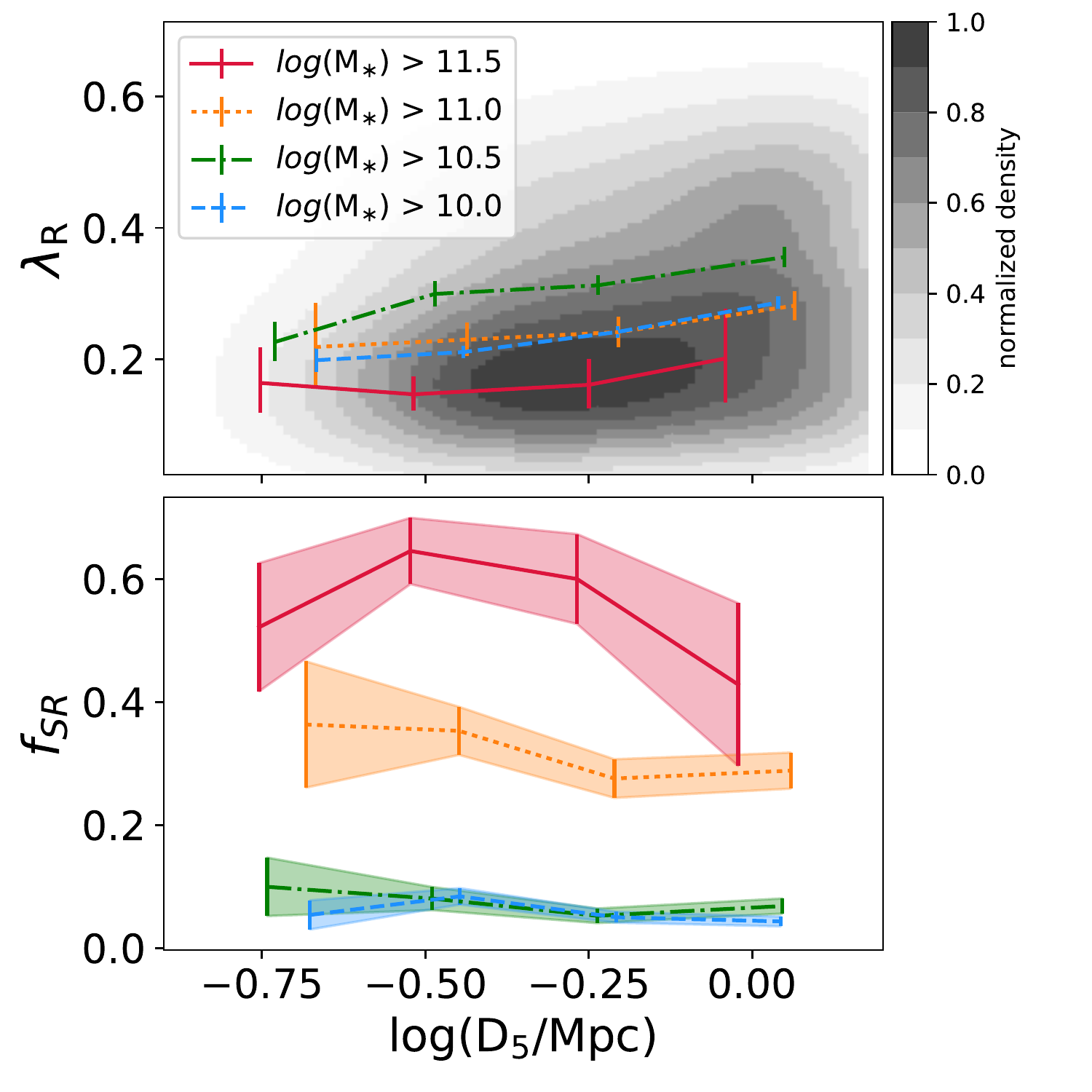}
\caption{Dependence of \lamR\ and $f_{\rm SR}$ on the environmental density D$_{5}$. 
 (A) Mean \lamR\ of each mass bin with error bars indicating 95\% confidence intervals of mean from bootstrap resampling on top of the \lamR\ distribution (gray shade).
 (B) Slow rotator fraction in the same bin and the same color scheme as the upper panel, with colored shading indicating the binomial error. 
Markers and error bars are at the mean D$_{5}$ in each bin. 
\label{fig:l_env_by_mass}}
\end{figure}

To disentangle the environmental dependence from the mass dependence, we further split the sample into 
four mass bins in Figure~\ref{fig:l_env_by_mass}.
A mild correlation between \lamR\ and D$_{5}$ is found, with the exception of the most massive bin (Figure~\ref{fig:l_env_by_mass}A). 
However, $f_{\rm SR}$ in each of the mass bin show no evidence for trends (Figure \ref{fig:l_env_by_mass}B); a result that is consistent with the findings of \citet{Brough2017a, Greene2017a}.
We argue that the discrepancy between the mean \lamR\ and $f_{\rm SR}$ demonstrates that $f_{\rm SR}$ is only sensitive to very slow-rotating ETGs, and fails to capture the entire picture.

\subsection{Environmental dependence by group classification}\label{sec:subgroup}

\begin{figure}[h!]
\figurenum{6}
\includegraphics[width=0.5\textwidth]{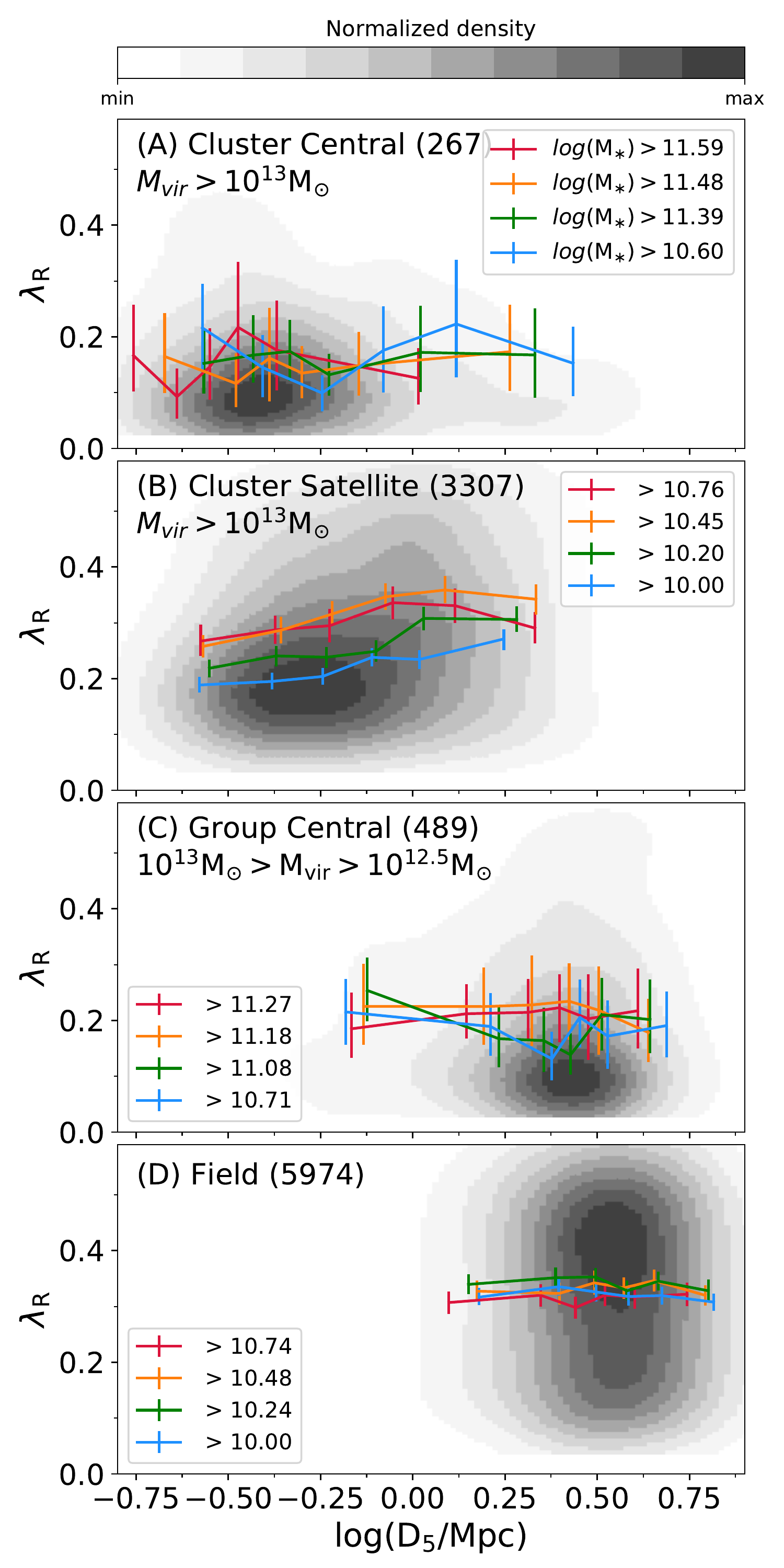}
\caption{Similar to the upper panel of Figure \ref{fig:l_env_by_mass} but for each population of ETGs.
In each panel, the samples are further divided into four subgroups of equal sample size.
The background number density map is normalized within each panel, and error bars indicate 95\% confidence intervals (CI) of mean from bootstrap resampling.
No trends are found in panels (A), (C) and (D), whereas a mild trend is found in panel (B).
\label{fig:l_env_pop}}
\end{figure}

In order to better understand the environmental dependence of spin, we split our ETG sample into four subsets: cluster central, cluster satellite, group central, and field ETGs. 
A group in this regard includes clusters, and thus  if the direct host halo of a galaxy exceeds $M_{\rm vir} > 10^{13} \thinspace \rm M_{\odot}$, we classify it as a cluster central.
Cluster satellites are satelite ETGs of halos with $M_{\rm vir} > 10^{13} \thinspace \rm M_{\odot}$.
If the host halo of a galaxy is within the mass range of $10^{12.5} \thinspace \rm M_{\odot} < M_{\rm vir} < 10^{13} \thinspace \rm M_{\odot}$, we categorize it as a group central galaxy.
Finally, the remaining galaxies (satellites of halos $M_{\rm vir} < 10^{13} \thinspace \rm M_{\odot}$ and centrals of halos $M_{\rm vir} < 10^{12.5} \thinspace \rm M_{\odot}$) are defined as field ETGs. 
Below the halo mass of $M_{\rm vir} < 10^{13} \thinspace \rm M_{\odot}$, satellites do not have a sufficient mass difference to the host halo because only the galaxies more massive than $M_{\ast} = 10^{10} \, \rm M_{\odot}$ are considered. There are 57 (0.95\%) such \textit{satellite} ETGs in the field ETG group.
Additionally, below the halo mass of $M_{\rm vir} < 10^{12.5} \, \rm M_{\odot}$, central galaxies are not dominated by mergers.
For this reason, we found this mass cut best represents the differential spin evolutions of central and satellite galaxies owing to their varying dependence on mergers and environmental effects.

We found 267 cluster centrals, 3,234 cluster satellites, 475 group centrals, and 5,721 field ETGs.
Note that, owing to our selection criterion of the ETG sample based on the color-magnitude diagram (See Figure~\ref{fig:sampling}), the field ETG sample may include a small fraction of LTGs. 
Thus, we consider field ETGs to represent a control sample of cluster satellites with similar properties except for their environments.

Figure \ref{fig:l_env_pop} shows the correlation between the mean \lamR\ and D$_{5}$ for each subsample according to the division of each group into four same-sized mass bins.
The cluster centrals are consistently rotating slowly throughout all environments (Figure \ref{fig:l_env_pop}A),
whereas the mean \lamR\ of cluster satellites in Figure \ref{fig:l_env_pop}B demonstrate decreasing trends toward denser environments in all mass bins,
especially in the dense region with D$_{5} < 1 \thinspace \rm Mpc$ (Figure \ref{fig:l_env_pop}B).
The group central ETGs comprise a combination of both high-mass, slow-rotating ETGs, and low-mass, fast-rotating ETGs (Figure \ref{fig:l_env_pop}C),
implying this is where the transitions of the central galaxies from FR to SR is operating.
Lastly, the field ETGs all exhibit fast rotation at mean \lamR\ $\approx 0.35$ regardless of their stellar masses or environments (Figure \ref{fig:l_env_pop}D).

When combined, cluster centrals and group centrals form a continuous distribution. 
They are similar in mass (group centrals are slightly less massive) and uniformly slow-rotating (group centrals rotate slightly faster).
The same is true for cluster satellites and field ETGs; both share similar mass ranges and their \lamR\ distribution are continuous. %, with decreasing \lamR\ only in dense environments.
Taken together, centrality appears to be a major factor in determining the spin of ETGs.

A number of other findings transpire from our measurements.
First, satellites have lower spins than field ETGs, with decreasing \lamR\ toward denser environments.
Second, smaller satellite ETGs have a lower spin. 
This trend is not visible in field ETGs, 
and is opposite to the trend observed for the entire ETG sample (see Figure \ref{fig:l_env_by_mass}).
These differences indicate that the factors responsible for spin loss likely differs from those operating to influence the mass growth of cluster satellites.

It is notable that cluster satellites alone (in Figure \ref{fig:l_env_pop}B) appear to show some environmental dependence in addition to the mass dependence.
We find that this is in reasonable agreement with the previous observational data of \cite{Brough2017a} based on a cursory inspection of their Figure 11.

\subsection{Varying Importance of Mergers}

\begin{figure*}[t]
\figurenum{7}
\includegraphics[width=0.95\textwidth]{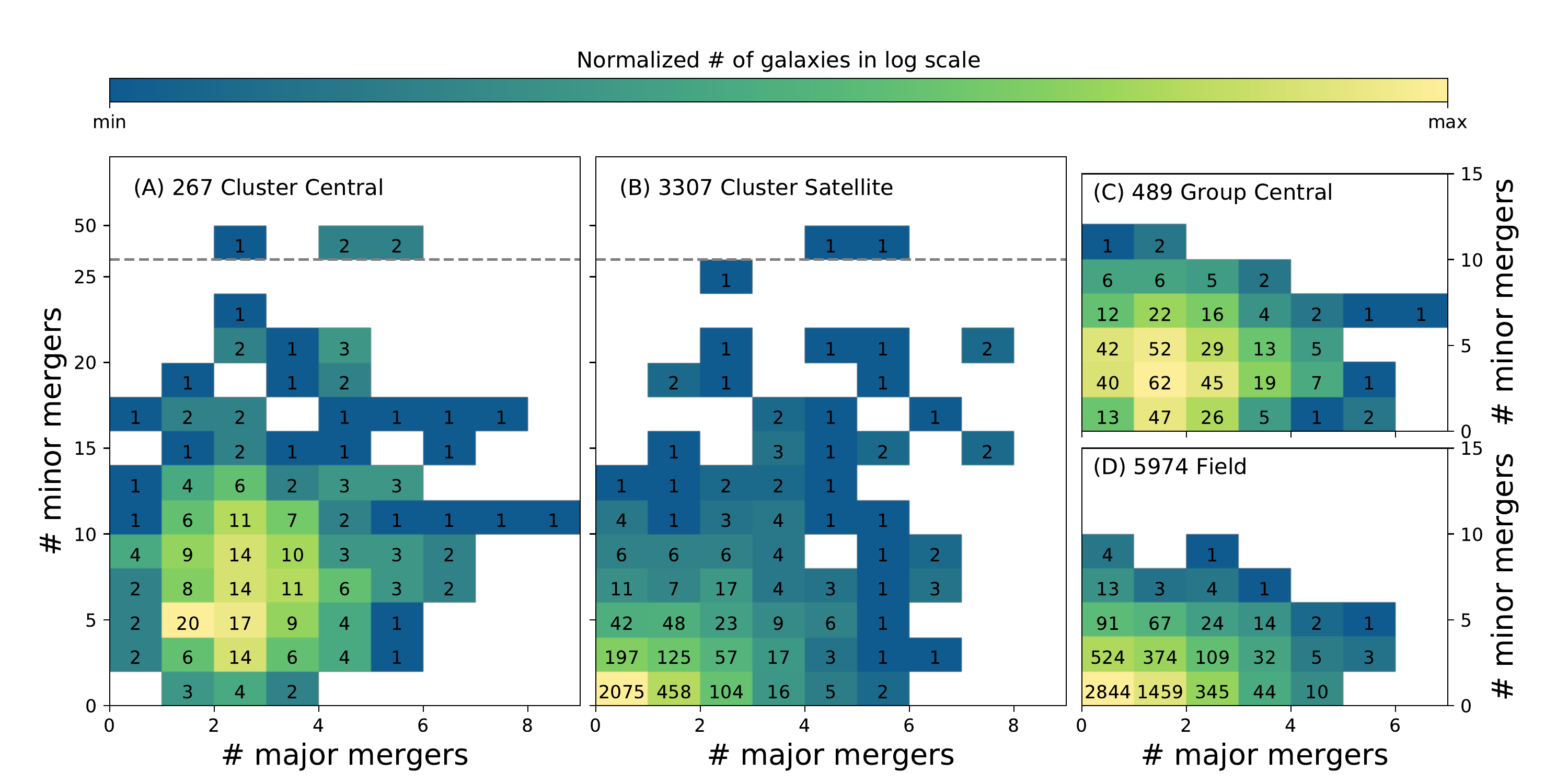}
\caption{Numbers of major merging (x-axis) and minor merging (y-axis) galaxies since $z=1$, color-coded based on the relative number of galaxies in each subgroup. 
Group central ETGs (A) have the broadest peak at around two - four major mergers and five - ten minor mergers.
By contrast, a large fraction of cluster satellite ETGs (B) undergo no major or minor mergers. The peak at no merger is the narrowest among the four groups.
Field central ETGs (C) form a peak at one major merger and two - four minor mergers. The peak is not at (0,0) merger, but the overall merger count is slightly less than the Group centrals. 
Field normal ETGs (D) have a peak at (0,0) almost as sharp as cluster satellites and a shorter tail along the $\rm N_{minor\ merger}$ direction compared to cluster satellites.
\label{fig:merger_freq}}
\end{figure*}

\begin{figure}[b]
\figurenum{8}
\includegraphics[width=0.5\textwidth]{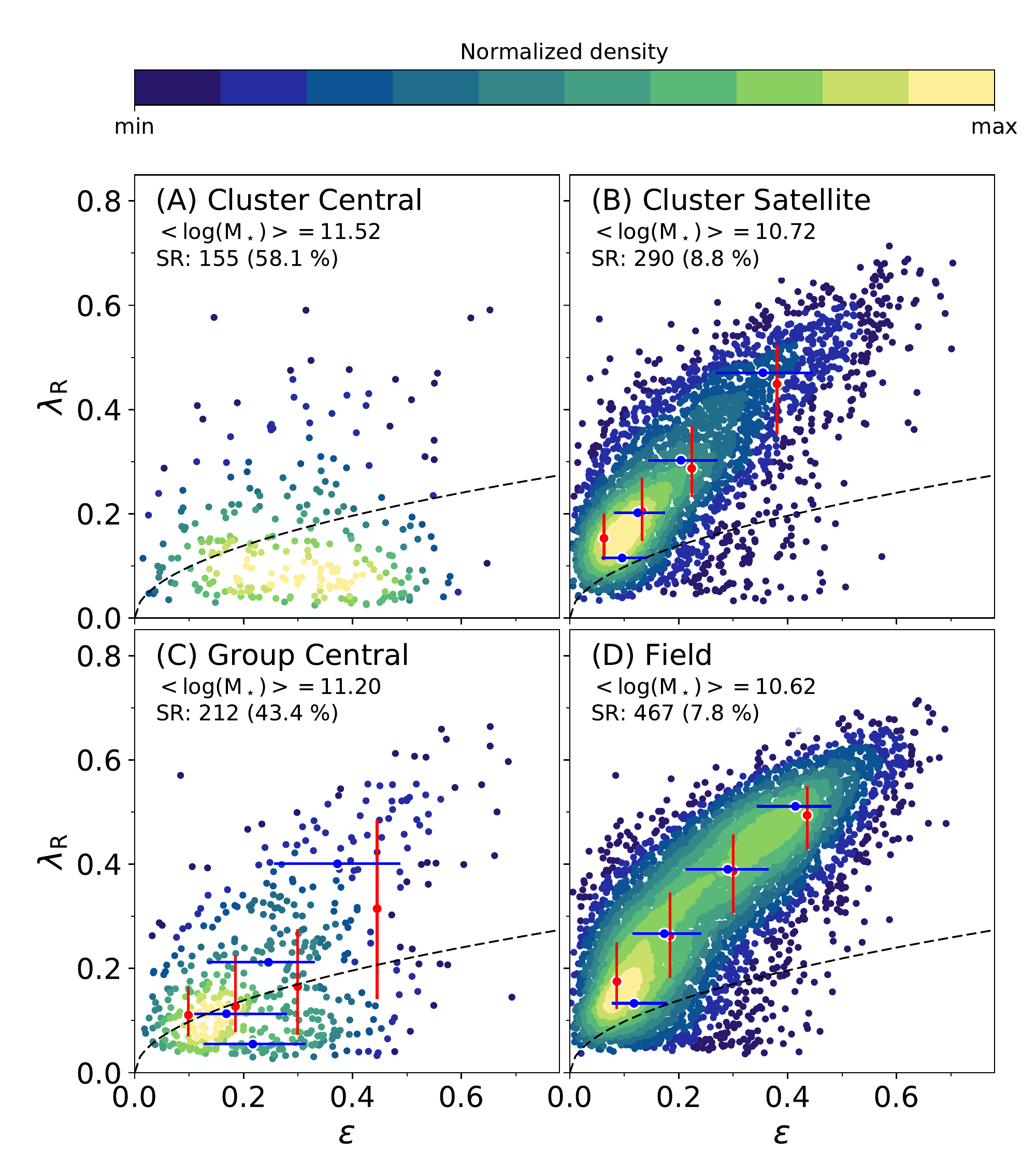}
\caption{Spin parameter \lamR\ versus ellipticity $\varepsilon$ as in Figure \ref{fig:l_e_mass} for the cluster central(A), cluster satellite(B), group central(C), and field satellite(D) galaxies.
Blue and red error bars represent $20$ -- $80\%$ range of $\varepsilon$ and \lamR\ independently binned along each axis.
\label{fig:l_e_pop}}
\end{figure}

In this section, we discuss the relative importance of mergers in each of the subgroups.
In Figure \ref{fig:merger_freq}, we show a two-dimensional histogram of galaxies with a given number of merger events since $\rm z = 1$.
Differences among merger counts of different subgroups are apparent.
A large fraction of cluster centrals have undergone two to five major mergers (mass ratio $>$ 1/4) and five to twelve minor mergers (Figure \ref{fig:merger_freq}A).
On the other hand, more than half (62.7\%; 2075 out of 3307) of the cluster satellites have had no major mergers, and at most one minor merger (Figure \ref{fig:merger_freq}B).
The majority (86.3\%; 2855 out of 3307) of the cluster satellites fall within the four (2 by 2) bins with the lowest merger counts.
Merger counts of the group centrals are between those of cluster centrals and cluster satellites (Figure \ref{fig:merger_freq}C).
In general, they undergo fewer mergers than cluster central ETGs, but considerably more than cluster satellite ETGs. 
Field normal ETGs are concentrated within low merger count bins, but not as much as cluster satellite ETGs (Figure \ref{fig:merger_freq}D). 
In summary, cluster centrals have the largest number of mergers among four subgroups, followed by group centrals, field ETGs, and then cluster satellite ETGs.

\begin{figure*}[ht]
\figurenum{9}
\includegraphics[width=0.95\textwidth]{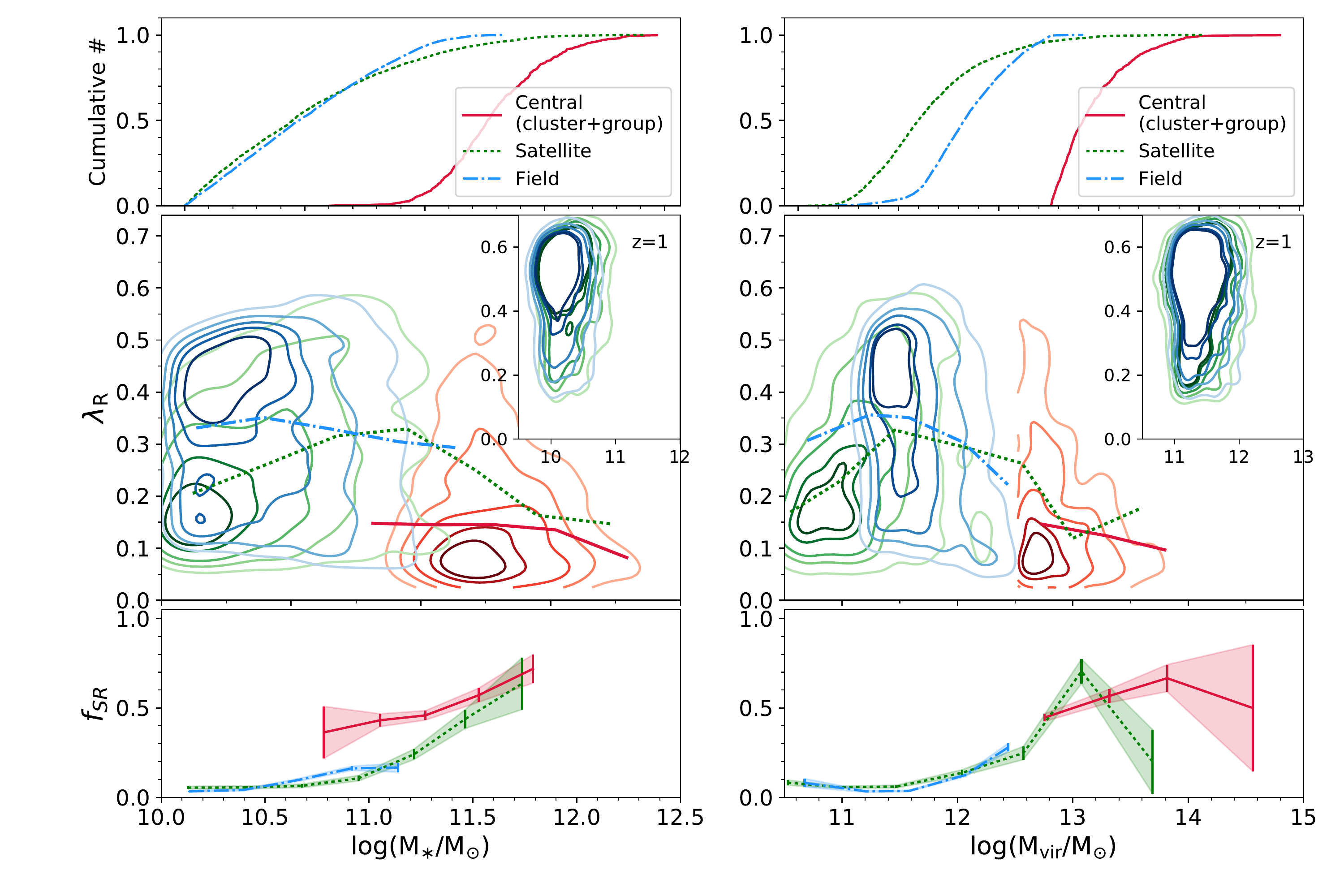}
\caption{(Top) Cumulative mass function of galaxies normalized for each subgroup. 
(Middle) \lamR\ dependency on stellar mass (left) and host halo mass (right), for each subgroup.
(Bottom) Slow rotator fraction $f_{\rm SR}$ plotted as a function of mass for each subgroup.
The mean galaxy spin \lamR\ of the central, satellite, and field ETGs is indicated by the solid red line, dotted green line, and dot-dashed blue line, respectively. 
Each contour line encompasses 10, 20, 40, 60, and 80$\%$ of each sample, respectively.
Note that cluster central and group central galaxies are combined in this plot for the purposes of clarity. 
Each inset panel shows the distributions of the progenitors of field and satellite galaxies at $z=1$ in the same color scheme, contour levels, and axes as the middle panel.
Because two subgroups are split by host halo masses, and their stellar masses roughly correlate with their host halo masses, it can be assumed that two-thirds of the central galaxies at the lower-mass end are group central ETGs.
\label{fig:l_vs_mass_pop}}
\end{figure*}

Given the dissimilar merger counts of subgroups, we now revisit  \lamR\ versus $\varepsilon$ in Figure \ref{fig:l_e_pop}.
Group centrals predominantly occupy the low \lamR\ region with a considerable scatter in $\varepsilon$ (Figure \ref{fig:l_e_pop}A). 
The peak of cluster satellite distribution is in a low \lamR\ region, but there is a significant fraction stretching toward higher \lamR\ and $\varepsilon$ region (Figure \ref{fig:l_e_pop}B).
The difference in $f_{\rm SR}$ is significant at 58.1\% (cluster centrals) versus 8.8\% (cluster satellites).
Finally, the group central ETGs show a larger scatter in the $\varepsilon$ direction and lower mean \lamR\ compared to the field ETGs (Figure \ref{fig:l_e_pop}C and D).

Comparing Figure \ref{fig:merger_freq} and Figure \ref{fig:l_e_pop}, a strong correlation between the number of mergers and the distribution in the $\lamR-\varepsilon$ plane is evident. 
First, the fact that most of the central galaxies have low spin and high $\varepsilon$ implies that 
mergers play an significant role in the kinematic evolution of cluster centrals after $z=1$ \citep[e.g.][]{Naab2014}.
By contrast, it is a rare occasion for cluster satellites to be disturbed by numerous mergers.
Second, a comparison between the \lamR\ distribution of cluster satellite and field ETGs reveals that the effect of a dense environment results in cluster satellite ETGs having a lower $\varepsilon$ with smaller scatter, as well as lower mean \lamR\, which is in the {\em opposite} direction of the merger effects (Figure \ref{fig:l_e_pop}B, D).

\subsection{Mass dependence by subgroup}
In Figure \ref{fig:l_vs_mass_pop} we revisit the galaxy spin divided in subgroups, focusing on the dependence on galaxy stellar mass and their host halo mass.
This time, we combined cluster centrals and group centrals together because they share the same domain in the \lamR\ versus stellar mass or host halo mass plane.

We found that the mean \lamR\ trend of satellite ETGs is bent (see also Figure~\ref{fig:l_e_mass_env}): 
less massive satellite ETGs rotate slower at the low-mass regime below $10^{11} \rm M_{\odot}$,
but more massive satellite ETGs rotate slower at the high-mass regime above $10^{11} \rm M_{\odot}$ (Figure \ref{fig:l_vs_mass_pop}, middle panel).
In the mass range of $10^{11} \thinspace \rm M_{\odot} < \rm M_{\ast} < 3 \times 10^{11} \thinspace \rm M_{\odot}$,
$\lamR$ as well as $f_{\rm SR}$ of the satellite differs from that of the central galaxies,
whereas they become similar above $3 \times 10^{11} \thinspace \rm M_{\odot}$.
The strong mass dependence of satellite ETGs at the high-mass end can be explained by the very large cluster satellites that are, concurrently, central galaxies of groups, i.e. due to mergers.

Comparing satellite ETGs and field ETGs at the low-mass regime reveals some interesting results.
The stellar mass distribution of field ETGs and satellite ETGs are similar, but the \lamR\ distributions are markedly different.
The mean \lamR\ of field ETGs is nearly constant over the mass range, but the mean \lamR\ of the satellite ETGs clearly declines with decreasing stellar mass.
The latter is directly the opposite of the negative mass dependence of spin of the most massive ETGs in groups, as previously seen from Figure \ref{fig:l_e_mass_env}.
Also note that $f_{\rm SR}$ of field ETGs and cluster satellite ETGs are almost identical and constant below $M_\ast\simeq 10^{11} \rm M_{\odot}$.
At $\left< \lambda_{\rm R} \right>\simeq 0.2$, cluster satellites at lower-mass end have a $f_{\rm SR}$ less than 0.1, 
whereas cluster satellites at $M_{\ast} \approx 2 \times 10^{11}$ have $f_{\rm SR} \approx 0.25$.
This implies that spin-down mechanism at the lower-mass end and higher-mass end are different.

Following the stellar mass dependence, galaxies embedded in sufficiently massive host halos ($\rm M_{\rm vir} > 10^{12.5}\,M_{\odot}$) 
have comparable spins, regardless of whether they are satellites or centrals (Figure \ref{fig:l_vs_mass_pop}, right panels).
For low-mass halos, the mean \lamR\ of satellites decreases with decreasing halo mass, whereas no clear trend is 
found in field ETGs. 
Although we have not included in this paper, we found that the halo mass distribution 
of field ETGs and cluster satellites is comparable at $z=1$, but the mass of satellite halos becomes smaller at $z=0$. 
This suggests that galaxy spin down and halo mass decrease occurred in conjunction.

We note that although two distinct mass dependencies are found in cluster satellites, 90\% of the population fall in the low-mass regime below $M_{\rm \ast} < 10^{11}$.
This implies that the environmental effects are influential for the spin evolution of the majority of satellite ETGs
and the merger-induced spin evolution, albeit stronger, is responsible for only small number of massive 
cluster satellite galaxies and central galaxies.

\begin{figure}[t!]
\figurenum{10}
\includegraphics[width=0.45\textwidth]{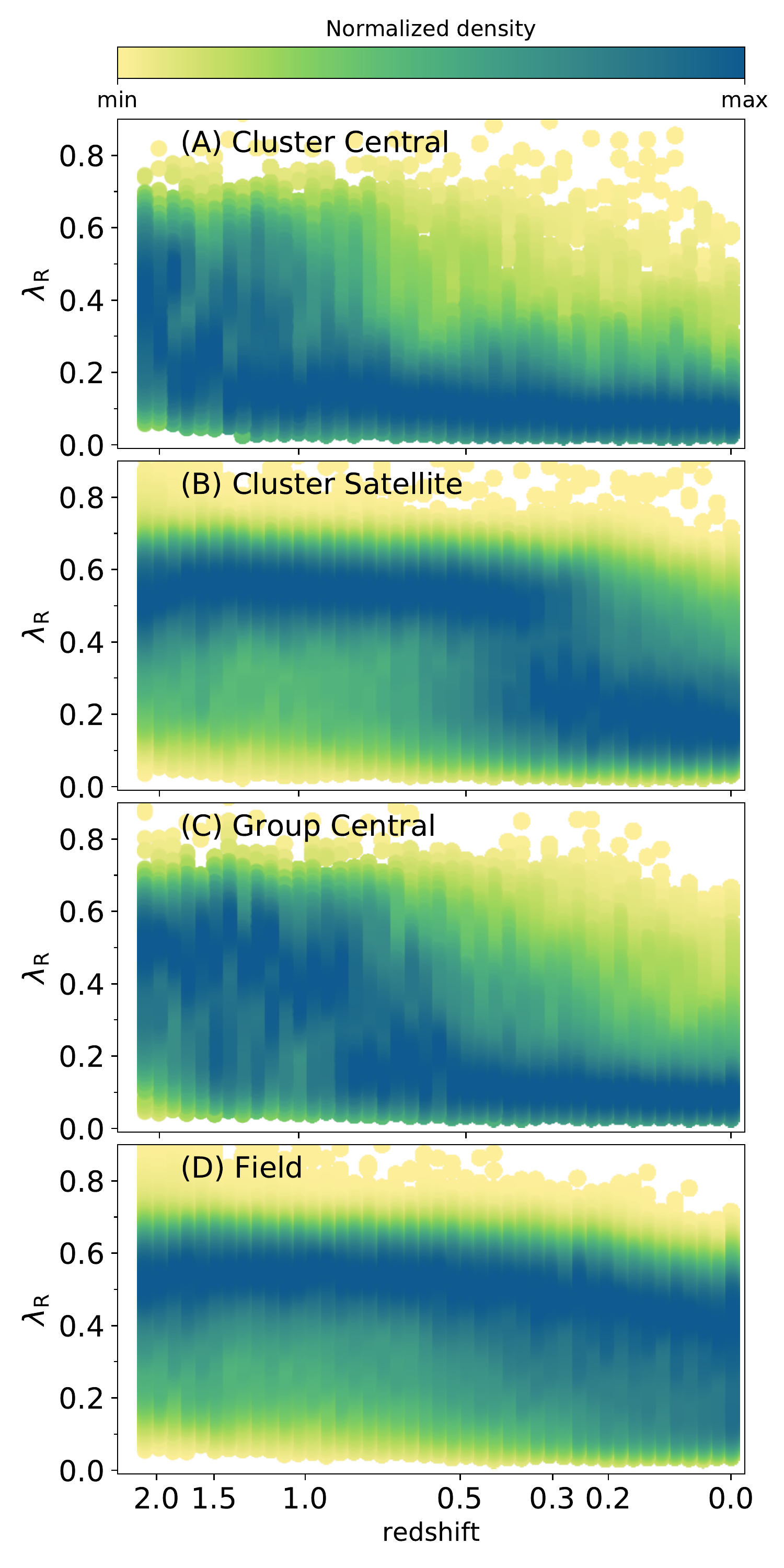}
\caption{Evolution of \lamR\ in subsamples of different environment since $z=2$.
We followed the main progenitor of each galaxy.
The x-axis, representing the redshift, is linear in physical time.
\label{fig:l_evol_pop}}
\end{figure}

\subsection{Evolution of Galaxy Spin}
Here, we  move on to the time evolution of \lamR.
In Figure \ref{fig:l_evol_pop}, we plot \lamR\ of sample galaxies and their main progenitors starting from $z=2$. 
We find that a significant fraction of cluster centrals at $z=2$ already rotate slowly (Figure \ref{fig:l_evol_pop}A).
By $z=1$, most of the cluster centrals rotate slowly at $\lambda_{\rm R} < 0.2$, and the density peak at low \lamR\ becomes more pronounced over time.
Although the main progenitors of cluster satellite, group central, and field ETGs all initially rotate fast with density peaks at $\lambda_{\rm R} \approx 0.6$, their evolution tracks differ substantially.

Figure \ref{fig:l_evol_pop}B illustrates a notable shift in the density peak of cluster satellite progenitors after around $z = 0.4$. %(It remains unclear why such a rapid shift occurs.)
Eventually, the density peak moves toward $\lambda_{\rm R} \approx 0.2$ with a broad tail toward higher \lamR.
The group central galaxies in Figure \ref{fig:l_evol_pop}C demonstrate noisy \lamR\ evolution track with a peak at a very low value of $\lambda_{\rm R} \sim 0.1$ appearing since $z=0.5$.
Without a merger or the effects from dense environments, field ETGs largely maintain their spin over time. 
Only a mild and gradual decline can been observed since $z=0.5$.
Despite the fact that cluster satellites and field ETGs show comparable mean stellar mass and \lamR\ distribution at $z=2$, their \lamR\ evolution tracks are markedly dissimilar.
Considering that cluster satellites undergo even fewer mergers than field ETGs, the larger spin-down of cluster satellites represents the importance of their environment, not necessarily mergers.

\begin{figure}[ht]
\figurenum{11}
\includegraphics[width=0.45\textwidth]{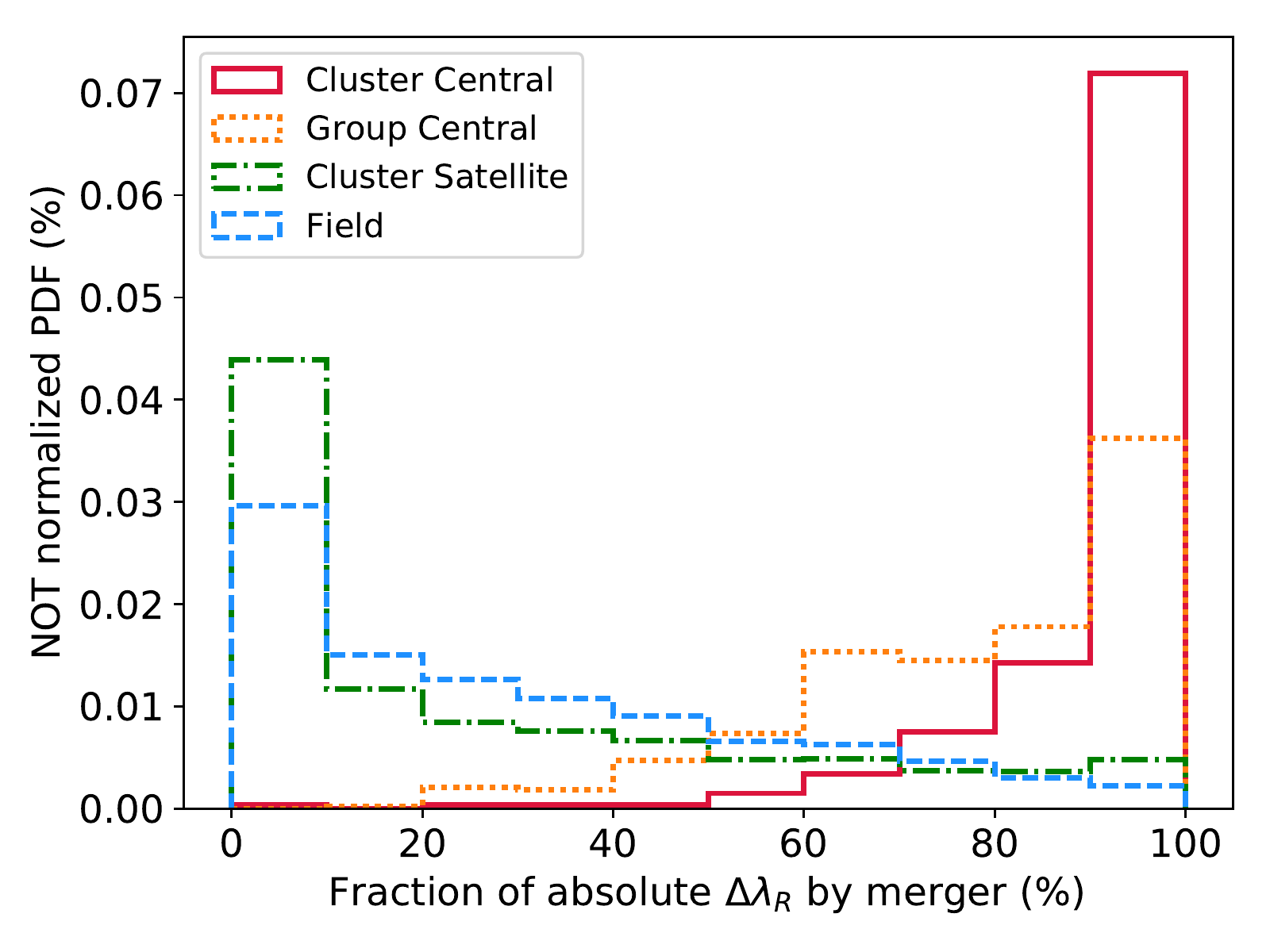}
\caption{Histogram of the fraction of absolute \lamR\ change between snapshots occurred during merger events since $z=2$.
Note that a galaxy with a 100\% fraction means that it is continuously affected by one or more mergers since $z=2$.
We show the absolute value to highlight the ability of mergers to make changes in galaxy spin in any direction.
\label{fig:Merger_importance}}
\end{figure}

Next, we present the relative importance of mergers in galaxy spin evolution. 
Specifically, we measured the absolute \lamR\ change between adjacent snapshots 
and determined if there were merger events taking place simultaneously. 
If a galaxy was affected by one or more mergers continuously since $z=2.0$, 
we determined the merger contribution to the spin change to be 100\%.
However, this does not necessarily imply that the physical origin of the \lamR\ change is entirely the results of mergers because other concurrent mechanisms could be operating.
Hence the results refer only to the upper limit of the merger impact.
Even as an upper limit, the majority of the satellite galaxies are only marginally 
influenced by merger events, mainly due to the lack of galaxy mergers (Figure~\ref{fig:Merger_importance}). 
In contrast, as discussed in the previous section, mergers appear to be largely responsible for the spin 
change in the cluster central and group central ETGs.

\subsection{Non-merger tidal interactions}
The gravitational potential of cluster/group environments leads to low probabilities of galaxy mergers, but a higher likelihood of high-speed encounters.
Hydrodynamical processes, such as ram pressure stripping or strangulation, are also important aspects of environmental effects in clusters; however, they do not seem to directly perturb the stellar structure \citep[e.g.][]{Smith2012}, and can only passively influence the galaxy properties by removing cold gas and shutting down new star formation.
Although forming new stars from late accretion is a possible way to spin up ETGs \citep{Lagos2017a, Penoyre2017}, star formation is insignificant in both cluster central and field ETG samples, and thus we
focused on tidal effects in our analysis.

To quantify the cumulative tidal effects on each satellite galaxy through non-merger encounters, we used the perturbation index (PI) inspired by \citet{Byrd1990}, which is defined as:
\begin{equation}\label{eq:PI}
{\rm PI} = \log { \left[ \int _{ t_{0} }^{ t }{ \sum _{ i=0 }^{ n }{ \left( \frac { { M }_{ p,i } }{ { M }_{ gal } }  \right) \times { \left( \frac { { R }_{ gal } }{ { d }_{ p,i } }  \right)  }^{ 3 }dt }  } \bigg/ Gyr  \right]  }, 
\end{equation}
where $M_{\rm gal}$ and $R_{\rm gal}$ are the mass and size of the galaxy in question,
$\rm M_{p,i}$ and $d_{p,i}$ are the mass and distance to the i-th perturber, respectively, and 
the time integration runs between two epochs in Gyr unit.

Any galaxy exceeding $5 \times 10^{8} \thinspace \rm M_{\odot}$ within 2$\thinspace \rm h^{-1}$Mpc from the galaxy in question is incorporated into the summation. 
Thanks to the power of -3 dependence on the distance, the PI quickly converges at a scale of few hundreds of kpc.

The PI is primarily designed to measure the tidal effect of the global potential field, but is also sensitive to galaxy harassment and galaxy mergers (one major merger corresponds to PI $\approx \log 1= 0$).
Thus, in this and the following sections, we focus our discussion on galaxies that have undergone no major or minor mergers since $z=1$.
This strict criterion leaves $47\%$ (1520) of cluster satellites, and $31\%$ (1767) of field ETGs. 
We checked that including galaxies with up to three minor mergers (leaving $69 \%$ and $56\%$ of each group, respectively) in the analysis did not affect the main results.

\begin{figure}
\figurenum{12}
\includegraphics[width=0.45\textwidth]{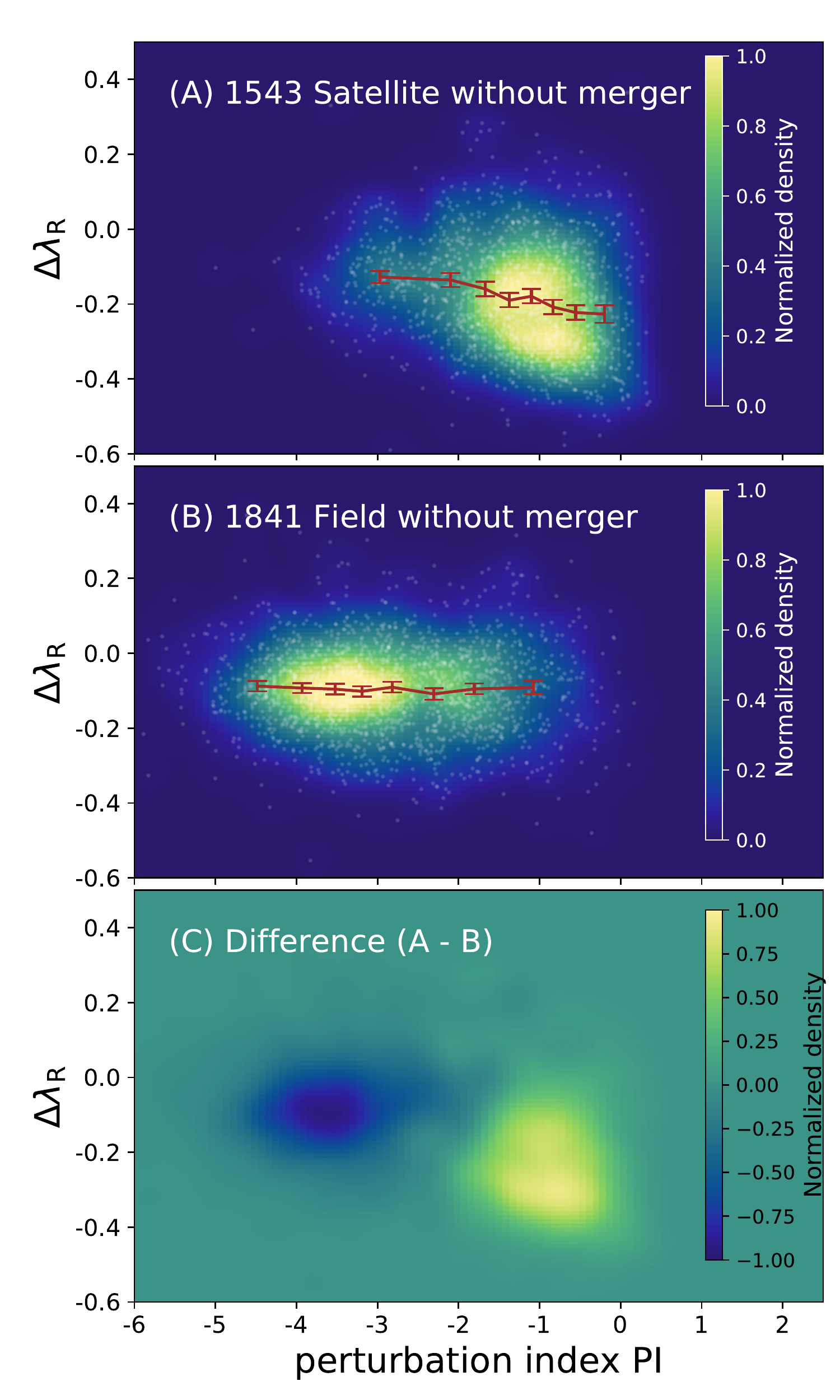}
\caption{Change in \lamR\ from $z=1$ to $z=0$ of satellite galaxies (A), and field ETGs (B), 
and the difference between the two populations (C). 
Only galaxies without mergers are considered in this plot.
The color scale is identical across all three panels; representing normalized height according to the peak in panel (A).
A considerable fraction of field ETGs have undergone  negligible tidal interactions, 
whereas a large fraction of satellites are severely affected by tidal interactions and lose their spin by more than $\Delta \lambda_{\rm R} \simeq -0.2$.
Red error bars indicate 95\% confidence intervals of mean from bootstrap resampling.
\label{fig:dl_dp}}
\end{figure}

In Figure \ref{fig:dl_dp}, we compared the amount of spin change against the PI of cluster satellite and field ETGs without mergers since $z = 1$.
The cluster satellites form a peak at $\rm PI\simeq-1$ and $\Delta \lambda_{\rm R} \simeq -0.25$ in a triangular-shaped envelop (Figure \ref{fig:dl_dp}A).
Although the scatter in $\Delta \lambda_{\rm R}$ intensifies with an increasing PI, galaxies with a larger PI show a larger drop in \lamR\ since $z=1$. 
It is also interesting to see that most galaxies with a low PI ($\sim-3$) are clustered around a small $\Delta \lambda_{\rm R}$ ($ \sim -0.1$), meaning that it is difficult for a galaxy to lose its spin significantly without being affected by either mergers or environmental tidal effects.
By contrast, field ETGs are only minimally perturbed ($\rm PI \la -2$) (Figure \ref{fig:dl_dp}B).
This indicates that the distributions of cluster satellite and field ETGs are almost mutually exclusive (Figure~\ref{fig:dl_dp}C).
Given that the mean stellar mass, mean \lamR\, and mean sSFR of satellite galaxies and field ETGs are comparable at $z\simeq1$ (Figure \ref{fig:props}), the difference in their evolution suggests that environmental differences play a major role in determining the galaxy spin in the absence of mergers.

\subsection{A scenario of satellite ETG evolution}\label{sec:scenario}

\begin{figure}
\figurenum{13}
\includegraphics[width=0.42\textwidth]{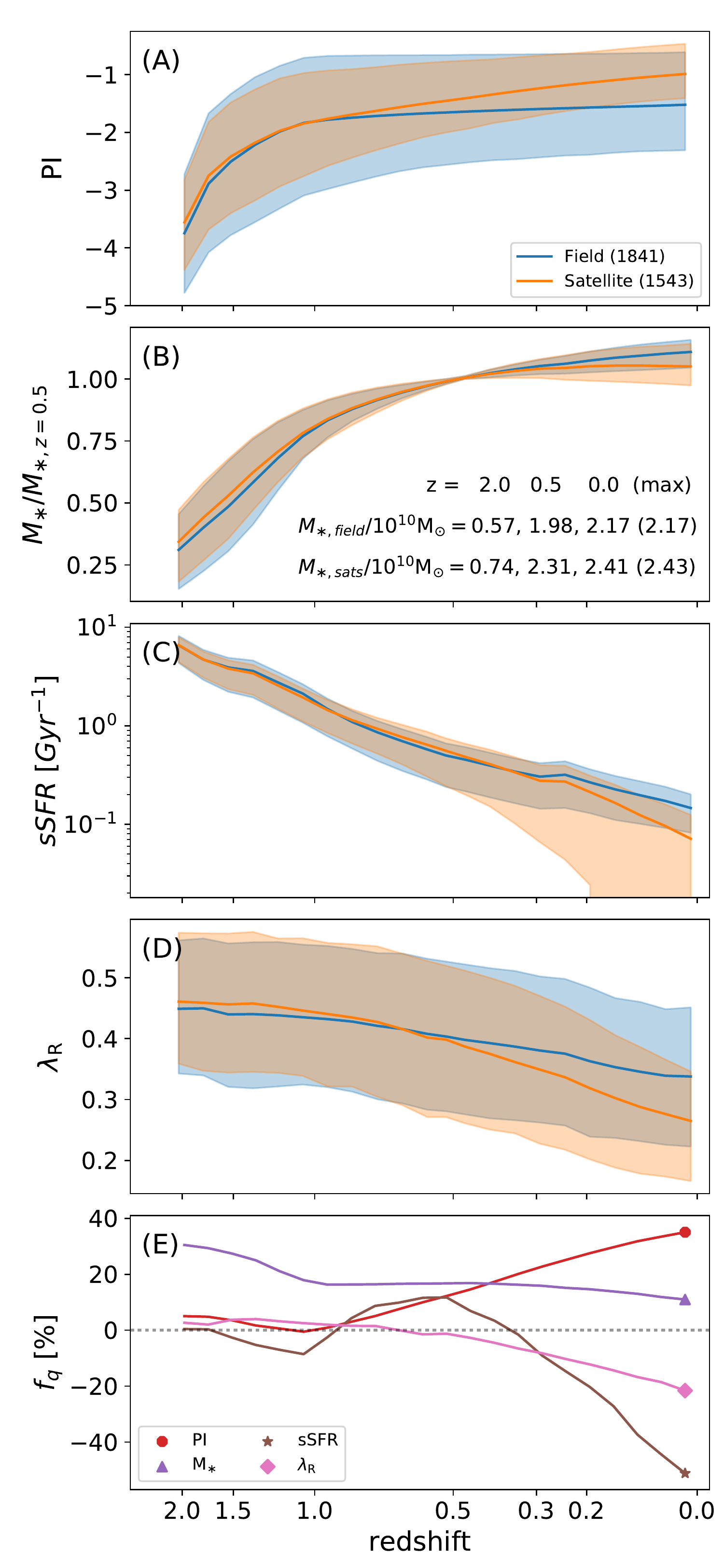}
\caption{Evolutionary trajectories of mean properties of 1,543 cluster satellites and 1,841 field counterparts without mergers since $z=1$.
(A) Perturbation index. 
(B) Fractional change in stellar mass with respect to the stellar mass at $z = 0.5$.
(C) Specific star formation rate. 
(D) Galaxy spin.
(E) Fractional residuals of cluster satellite properties with respect to those of field ETGs.
 Shaded regions indicate the upper and lower quartiles ($25-75\%$) of each distribution, and solid lines represent the mean values.
The x-axis, representing the redshift, is linear in physical time. 
Note that the sudden drop of the lower envelop occurs when the bottom 25\% of the satellites have an undetectable star formation (no new stellar particle).
In panel (E), unlike in panel (B), we plot the fraction of the difference in stellar mass ($ [M_{*,\rm GS}-M_{*,\rm FN}]/M_{*,\rm FN}$).
\label{fig:props}}
\end{figure}

A more thorough picture of the spin evolutions of cluster satellite ETGs in comparison with field counterparts is presented in Figure \ref{fig:props}.
Again, we only consider galaxies with no merger events since $z=1$ for a more robust comparison.
We plot the evolution of the mean PI of cluster satellites and field ETGs in orange and blue, respectively, and $25-75\%$ ranges through shadings.
Albeit the large scatter, the typical PI of cluster satellites and field ETGs is comparable in the early epoch ($z\sim2$), but they begin to deviate at $z \simeq 1$, as only cluster satellites are continually perturbed by interactions with nearby galaxies (Figure \ref{fig:props}A).
In other words, the environmental difference of the two populations becomes noticeable at $z \simeq 1$.

In Figure \ref{fig:props}B, we plot the fractional mass change of galaxies compared to the mass at $z=0.5$, after which the environmental effects accelerate the deviation in the properties of the cluster satellites and the field ETGs.
The mean stellar masses of the two populations at $z=2$ are $5.7 \times 10^{9}\thinspace \rm M_{\odot}$ and $7.4 \times 10^{9}\thinspace \rm M_{\odot}$, respectively.
Contrary to the persistent stellar mass growth of field ETGs until $z=0$, the stellar mass of cluster satellites grows more slowly and eventually decreases after $z=0.2$.
The trends in sSFR (Figure \ref{fig:props}C, measured over the previous 0.1 Gyr) and \lamR\ (Figure \ref{fig:props}D) are similar in that cluster satellite ETGs initially have comparable values as the field ETGs, but ultimately have lower values than field ETGs at $z=0$, probably due to ram pressure stripping of cold gas in the former case.

To highlight the differential evolution of cluster satellite and field ETGs, we plot the fractional residual of each quantity ($q$) in Figure \ref{fig:props}E, which is defined as $ f_{q} \equiv \left<q_{\rm GS}\right>/ \left<q_{\rm FN}\right> - 1$,
where GS and FN represent cluster satellites and field ETGs, respectively.
Note that we plot the residual fraction of absolute stellar mass ($=\left<M_{*,\rm GS}\right>/\left<M_{*,\rm FN}\right>-1$), 
instead of the fractional change in stellar mass as in Figure~\ref{fig:props}B for a valid comparison. 
The initial properties of cluster satellite and field ETGs are comparable, except for the $25\%$ difference in the mean stellar mass.
Yet different properties seem to evolve at different rates.
The PI of cluster satellites begin to rise  significantly at $z \simeq 1$, accompanied by the slow increase in stellar mass, followed by a reduction in \lamR\ after $ z \simeq 0.5$.

It is also worth noting that the decline in sSFR lags behind the rise or fall in the rest of the properties. 
This suggests that tidal perturbations initiate the spin-down of satellite ETGs, whereas star formation quenching due to 
environmental effects may help to maintain the slow rotation at the later stages.

\section{Discussion and conclusion} \label{sec:summary}
We analyzed ETGs more massive than $10^{10}M_{\odot}$ from the $(100\, h^{-1} \, {\rm Mpc})^3$ volume of the Horizon-AGN simulation.
The simulated ETGs reasonably reproduced the loci of \lamR\ versus $\varepsilon$ distribution from IFU observations \citep{Emsellem2007, Cappellari2011}.
We found an increase in $f_{\rm SR}$ along the increasing stellar mass within a mass range of above $\sim 5 \times 10^{10} \thinspace \rm M_{\odot}$, which is in agreement with recent observations by \citet{DEugenio2013, Houghton2012, Fogarty2014, Cappellari2016, Veale2017a, Brough2017a, Greene2017a}.
While the marginal correlation between the mean \lamR and the environmental density of the total ETG sample is consistent with the observations, 
varying dependencies of ETG spin on stellar mass and environments were found when ETGs are divided into subgroups.

Having evolved through numerous mergers, more massive (both group and field) central ETGs rotate more slowly, regardless of their environment. 
This clear mass dependence of the spin of central ETGs is consistent with the aforementioned IFU observations and can be explained by the merger-driven formation scenario \citep{Naab2014, Lagos2017a, Penoyre2017}.
Field normals, with fewer mergers than central ETGs, more or less have maintained their \lamR\ throughout the cosmic evolution, showing no dependence on stellar mass or environments.
Interestingly, satellite ETGs in massive halos rotate much slower than field ETGs at similar stellar mass even though satellites undergoes fewer mergers than their field counterparts.

We demonstrated the importance of environmental tidal effects by comparing satellite ETGs and field ETGs whose progenitors shared similar properties until $z\simeq1$.
The satellite ETGs that have suffered from a higher degree of tidal perturbation since $z\simeq1$ showed a larger drop in their spin.
By contrast, field ETGs accumulated only negligible amount of tidal effects and showed correspondingly small $\Delta \lambda_{\rm R}$, unless they have undergone mergers.
A clear separation is made between satellite ETGs and field ETGs without mergers on the PI (Equation~\ref{eq:PI}) versus $\Delta \lambda_{\rm R}$ plane, which is caused by the differential environments after $z\simeq1$. 

Consequently, the \lamR\ of the satellite ETGs show distinct features because their spin evolution, with the absence of mergers, is mainly determined by cumulative effect of tidal perturbation as quantified by PI. 
First, the mass dependence of \lamR\ of the satellite ETGs is not monotonic.
The mean \lamR\ of the satellite ETGs below $\rm M_{\ast} < 10^{11} \rm M_{\odot}$ has a positive mass dependence,
which is opposite to the negative dependence of central ETGs in dense environments.
This outcome is likely because, among satellites, smaller ETGs are more susceptible to their environments, and thus lose spin more easily. 
Second, a clear environmental dependence of the mean \lamR\ is visible: cluster satellites in dense regions have lower mean \lamR\ than those in less dense region. 

Our selection criterion of ETGs involves possible contamination from LTGs (Section \ref{sec:sampling}).
Even so, we argue that a precise distinction between relatively slow-rotating LTGs and relatively fast-rotating ETGs is unimportant in our analysis,
considering the fact that galaxy properties are continuous functions of galaxy spin \citep{Cappellari2016}.
Nonetheless, we confirmed that adopting an alternative criterion, for example, that of \citet{Schawinski2014}, did not change our main conclusions.

Galaxy mergers are indeed prevalent at high redshifts ($z \gtrsim 2$),
but at lower redshifts they are the dominant driver of the spin down only for central ETGs in dense regions, which represent less than 10\% of the whole ETGs population. 
By contrast, environmental perturbation drives the slower rotation of cluster satellites compared to their field counterparts.
The details of our model prediction is at the moment more than challenging to test through current observational techniques.
Nonetheless, we expect that future observations using MUSE \citep{Bacon2010} or Hector \citep{Bryant2016} may be able to test our scenario by constructing a more extensive catalog over a wider stellar mass range.

\acknowledgments
SKY, acted as the corresponding author, acknowledges support from the Korean National Research Foundation (NRF-2017R1A2A1A05001116).
This study was performed under the umbrella of the joint collaboration between Yonsei University Observatory and the Korea Astronomy and Space Science Institute.
The analysis were carried on the Horizon Cluster hosted by Institut d’Astrophysique de Paris. 
We thank S. Rouberol for running it smoothly for us. 
This research is part of Spin(e) (ANR-13-BS05-0005, http://cosmicorigin.org).

\bibliography{Rotators2}
\end{document}